\documentstyle[12pt]{article}
\def\thefootnote{\fnsymbol{footnote}}
\begin{document}
%
\font\elevenbf=cmbx10 scaled\magstep 1
\renewcommand{\thefootnote}{\fnsymbol{footnote}}
\parindent=3pc
\baselineskip=10pt
\newcommand{\R}{\mbox{$R$}}
\newcommand{\Rb}{\mbox{$\not \! \! \R$}}
\newcommand{\Lb}{\mbox{$L$}}
\newcommand{\Lbb}{\mbox{$\not \! \! \Lb$}}
\newcommand{\B}{\mbox{$B$}}
\newcommand{\Bb}{\mbox{$\not \! \! \B$}}
\newcommand{\CP}{\mbox{$CP$}}
\newcommand{\nCP}{\mbox{$/ \! \! \! \! \! \! \rm CP$}}
\newcommand{\be}{\begin{equation}}
\newcommand{\ene}{\end{equation}}
\newcommand{\een}{\end{subequations}}
\newcommand{\ben}{\begin{subequations}}
\newcommand{\beq}{\begin{eqnarray}}
\newcommand{\eeq}{\end{eqnarray}}
\begin{center}
\hfill{FTUV/94-42, IFIC/94-37}\\
\hfill{PRL-TH/94-23}
\vskip 2.0truecm
{\Large \bf JUST SO OSCILLATIONS IN SUPERSYMMETRIC\\[.25cm]STANDARD MODEL
}
\vskip 1.0cm
{\large \bf ANJAN S. JOSHIPURA}
\footnote{On leave from: Theoretical Physics Group,
 Physical Research Lab., Ahmedabad, India},
\footnote{e-mail: joshipur@vm.ci.uv.es}
\vskip .5cm
{\it Instituto de F\'{\i}sica Corpuscular - C.S.I.C.\\
Departament de F\'{\i}sica Te\`orica, Universitat de Val\`encia\\
46100 Burjassot, Val\`encia, SPAIN}\\
\vskip 0.5cm
{\large \bf MAREK NOWAKOWSKI}
\footnote{Feodor-Lynen Fellow}\footnote{Present Address
:Laboratori Nazionali di Frascati, C.P.
13-Frascati (Rome), Italy}

\vskip .5cm
{\it Theoretical Physics Group,
Physical Research Laboratory, \\
Navarangpura, Ahmedabad, 380 009,INDIA}
\end{center}
\begin{abstract}
We analyze the spectrum and mixing among neutrinos in the minimal
supersymmetric standard model with explicit breaking of R parity.
It is shown that ({\em i}) the mixing among neutrinos is naturally
large and ({\em ii}) the non zero neutrino mass is constrained to be
$\leq 10^{-5}$ eV from arguments based on baryogenesis. Thus
vacuum oscillations of neutrinos in this model may offer a solution of the
solar
neutrino problem. The allowed space of the supersymmetric parameters
consistent with this solution is determined.
\end{abstract}
\newpage
\section {Introduction.}

The neutrino masses \cite{one} are known to solve some of the outstanding
problems notably, the solar neutrino , the atmospheric
neutrino  and the dark matter problem. Theoretically, the
presence of a non-zero neutrino mass provides a window into
physics beyond the standard electroweak model. The generation of
neutrino masses is possible either in the presence of neutral
Higgs transforming as an $SU(2)_L$-triplet and /or if there
exist additional neutral fermions with which the conventional neutrinos
could mix. The most popular example of the latter kind is
provided by the seesaw mechanism \cite{one} in which the left-handed
neutrinos obtain their masses through mixing with the
right-handed neutrinos. Another example is provided in
supersymmetric theory \cite{2} which automatically contains additional
neutral fermions, namely gauginos and Higgsinos. However, in this
theory neutrinos cannot mix with the latters if the Lagrangian
possess a symmetry, called $R$-parity ($R$) \cite{3} which distinguishes
between matter and supermatter. But breakdown of $R$-parity can
lead to mixing of neutrinos with gauginos and Higgsinos \cite{4,5}
and hence to mass of neutrinos. In fact, as long as the terms
associated with $R$-parity breaking are small, natural seesaw mechanism
is operative, with gauginos and Higgsinos playing the role of the
`right-handed neutrino' of the conventional seesaw mechanism.
The possibility of generation of non-zero neutrino mass in SUSY
models with broken $R$-parity is extensively discussed in the
literature, both in the Supersymmetric Standard Model
(SSM) \cite{5,6} as well as in some of its extensions \cite{7}. The
neutrino masses have in fact been used to put constraints on the
amount of admissible violation of $R$ \cite{5,6,7,8}. There exist independent
constraints on the strength of $R$-breaking. They arise by
requiring \cite{9} that the $(B-L)$ violation associated with
$R$-breaking should not erase the baryon asymmetry in the
presence of the sphelaron induced $(B+L)$ violation \cite{10}. This can
impose severe restrictions on the neutrino masses. In this paper
we wish to systematically analyze the structure of neutrino
masses and mixing in SSM with explicitly broken $R$-parity
utilizing the restrictions imposed on $R$-parity violating interactions.

The neutrino mass can offer a solution \cite{one} to the solar neutrino
problem if the relevant ( mass)$^2$ difference $\Delta$
lies either around $10^{-5}-10^{-6}\;eV^2$ or around $10^{-10}\;
eV^2$. The matter induced resonant oscillations \cite{11} deplete the
neutrino flux in the former while vacuum or `just so' oscillations
\cite{12} are responsible for the depletion in the latter case.
Moreover, the vacuum solution is feasible only if the mixing
angle $\theta$ between oscillating neutrinos is large, typically
$\sin^2 (2\theta) \sim 0.75-1$ \cite{one}.
Such solution therefore require extremely tiny neutrino masses which can
arise for example, from the Planck scale physics \cite{Planck}.
We shall show that the constraints
coming from baryogensis in SSM in fact restrict the neutrino mass to be
as small as $\sim 10^{-5}\; eV$ a value just required in order to solve the
 solar
 neutrino problem.
 The large mixing required  for this purpose
also
follows naturally in the SSM with broken $R$-parity, as we
will see. While restrictions on R parity breaking coming from
the baryogenesis are well known, their implications for
obtaining the vacuum solution to the solar neutrino problem have
not been stressed before. The purpose of the present note is to
point out this possibility and at the same time determine the
allowed region of parameters which realize it.

We discuss the SSM with $R$-parity violation and
summarize the constraints coming from baryogenesis in the next
section. In section 3 we discuss the structure of neutralino
masses in the presence of $R$-parity violation. An effective
seesaw mechanism allows one to reduce the $7 \times 7$ mass
matrix to a $3 \times 3$ effective neutrino mass matrix and
makes it possible to discuss the mixing among neutrinos
analytically. We determine in this section restrictions imposed
on the conventional parameters of SSM if one wants to solve the
solar neutrino problem. The $R$ violation causes the neutralinos
to decay into $Z^*$ and neutrino. We determine this coupling and
discuss its implications in section 4. Conclusions are presented
in the last section.

\section{ $R$-parity violating SSM.}

We shall confine ourselves to SSM. This is characterized by the
following superpotential in standard notation
\be \label{e1}
W_0 = \varepsilon_{ab}\left [h_{ij}\hat{L}_i^a \hat{H}_1^b
\hat{E}_j^C + h'_{ij}\hat{Q}_i^a \hat{H}_1^b \hat{D}_j^C + h''_{ij}
\hat{Q}_i^a \hat{H}_2^b \hat{U}_j^C + \mu \hat{H}_1^a \hat{H}_2^b \right]
\ene
In addition to $SU(2)_L \otimes U(1)_Y$, this potential is also
invariant under the discrete $R$-parity under which quarks,
leptons, Higgs and gauge bosons are even while their
superpartners are odd. $R$ can be broken in SSM explicitly by
the following terms
\be \label{e3}
W_{\Rb}=\varepsilon_{ab}\left[\lambda_{ijk}\hat{L}_i^a \hat{L}_j^b
\hat{E}_k^C + \lambda'_{ijk}\hat{L}_i^a \hat{Q}_j^b \hat{D}_k^C
+ \epsilon_i \hat{L}_i^a \hat{H}_2^b \right] + \lambda''_{ijk}
\hat{U}_i^C \hat{D}_j^C \hat{D}_k^C \ene
The presence of, both lepton and baryon number violating terms
in eq. (2) leads to difficulties with proton lifetime. We
therefore set from now on the coupling $\lambda''_{ijk}$ to zero.
The full superpotential $W$ is now the sum of $W_{0}$ and $W_{\Rb}$

Even in the absence of $R$-parity violating terms $W_{\Rb}$, $R$ can
be broken spontaneously if the sneutrino acquires a non-zero
vacuum expectation value (vev) $\tilde{\nu}_L$ \cite{13}.
This possibility is allowed in
SSM \cite{14}, but the resulting vev of the sneutrino field is large,
typically around weak scale for natural range of parameters. Moreover,
one generates a majoron which is strongly coupled to
$Z$. Such a majoron is in conflict with the invisible width of
the $Z$. It is possible to avoid this conflict by introducing a
small explicit R breaking term \cite{5,14,15}
$\varepsilon_{ab}\epsilon_i \hat{L}_i^a \hat{H}_2^b$ in $W$.
Then even for very small values of $\epsilon_i (\sim MeV)$, the majoron
can be made massive $(\sim 100\; GeV)$ \cite{15}. The vev of the sneutrino
is nevertheless  constrained ($< 5\;GeV$) from other
observations at LEP and from neutrino masses \cite{16}. Thus one must
abandon the idea of spontaneous R violation in the SSM.
 Restrictions on parameters of SSM
implied by this requirement were worked out in \cite{15}. We shall
assume these parameters to lie in the range determined in \cite{15}
and hence $<\tilde{\nu}_L>$ would be assumed zero in the absence
of $R$-breaking terms. But now if one introduces a small explicit
R breaking term $\varepsilon_{ab}\epsilon_i \hat{L}_i^a \hat{H}_2^b$
then the sneutrino vev automatically gets generated \cite{5,asj}. In this case,
the vev is  related to the explicit rather than to the spontaneous violation of
R parity and hence is not  accompanied by a massless majoron.
Moreover, as long as the R breaking parameters are small, the
sneutrino vev also remain small and one avoids conflict with
phenomenology which require the sneutrino vev to be small
independent of the existence of the majoron \cite{16}.
On the other hand the R breaking induced by $\epsilon_i$ and
the consequent sneutrino vev can lead to
 interesting predictions for neutrino masses.

The scalar potential following from $W$ and general soft supersymmetric
breaking terms has the following form \cite{asj}
\beq \label{e3p}
V_{Higgs}&=&\mu_1^2\vert \phi_1 \vert^2 + \mu_2^2 \vert \phi_2 \vert^2
+\mu_{L_i}^2 (\varphi_i^{\dagger}\varphi_i) \nonumber \\
&+& {1 \over 2}\lambda_1\left[\vert \phi_1 \vert^4 + \vert \phi_2
\vert^4 + (\varphi_i^{\dagger} \varphi_i)^2 + 2\vert \phi_1 \vert^2
(\varphi_i^{\dagger} \varphi_i) -2\vert \phi_2 \vert^2(\varphi_i^{\dagger}
\varphi_i) \right] \nonumber \\
&+&\lambda_2 \vert \phi_1 \vert^2 \vert \phi_2 \vert^2 -(\lambda_1
+\lambda_2)\vert \phi_1^{\dagger} \phi_2 \vert^2 +\left (\lambda_3
(\phi_1^{\dagger} \phi_2) + h.c. \right) \nonumber \\
&+&\left(i\kappa_i(\phi_1^T\tau_2 \varphi_i) +h.c. \right)
+ \left(i\kappa'_i(\phi_2^T \tau_2 \varphi_i) +h.c.
\right)+\cdots \eeq
where $\varphi_i \equiv L_i$, $\phi_2
\equiv H_2$ and $\phi_1 \equiv -i\tau_2 H_1^*$ and the dots
indicate terms of the potential not relevant for minimization.
In deriving eq. (3) we have assumed for simplicity that all
$\epsilon_i$ are equal. Then the parameters of the Higgs
potential are (we will neglect all possible CP-violating phases)
\beq \label{e4}
&&\mu_1^2=m_1^2+\vert \mu \vert^2,\; \; \; \mu_2^2=m_2^2 +\vert
\mu \vert^2 + \epsilon_i \epsilon_i, \; \; \; \mu_{L_i}^2=m_{L_i}^2+
\vert \epsilon_i \vert^2 \nonumber \\
&&\lambda_1={1 \over 4}(g^2 + g'^2), \; \; \; \lambda_2={1 \over
2}g^2- \lambda_1,\; \; \;  \lambda_3=-m_{12}^2, \; \; \;
\kappa_i=\mu \epsilon_i \eeq
The parameters $m_i^2$, $m_{12}^2$, $m_{L_i}^2$ and $\kappa'_i$
are soft breaking parameters. $\kappa'_i$ gets related to $\epsilon_i$
at Planck scale in the SSM. The $g$ and $g'$ are the gauge coupligs.

Because of the presence of the $\kappa_i$ and $\kappa'_i$ terms
in (3) the minimization invariably leads to non-zero vev
$\omega_i \equiv <\tilde{\nu}_{iL}>$ of the sneutrino. These are
given by \cite{asj}
\be \label{e5}
\omega_i\;=\;{\kappa_i v_1\; +\; \kappa'_i v_2 \over \mu_{L_i}^2
\;+\; {1 \over 2}\lambda_1 \left(\vert v_1 \vert^2\; -\; \vert v_2 \vert^2
\; + \; \sum_k \vert \omega_k \vert^2 \right)} \ene
We are working here in the unconventional basis in
which the $\epsilon$ term in $W$ is  not rotated away. Even if
one choses to utilize this freedom of rotation, the essential
ingredients remain unchanged. In particular, the vev for the
sneutrino gets generated independent \cite{5} of the basis one chooses.
Note from eq.(5) that $\omega_i$ vanish when the R breaking terms
$\kappa_i,\kappa_i'$ and $\epsilon_i$ are taken zero.
 In this limit the model reduce to the minimal standard model
 and thus has two scalars and a massive pseudoscalar.
 The presence of $\kappa,\kappa'$ and $\epsilon$ adds a small correction
 to these masses. Thus the spectrum of the model
does not contain a majoron quite in contrast with models with spontaneous
$R$ violation \cite{14,15}.

As we will see in the next section, the parameters $\epsilon_i$
and $\omega_i$ determine the tree level neutrino masses and
mixing. We therefore summarize the restrictions \cite{9} on these
parameters which follow from  the baryogenesis \cite{10}. Lepton number
violation induced by $\epsilon_i, \kappa_i$ or
$\kappa'_i$ could erase the existing baryon (or $B-L$) asymmetry if
the $(B+L)$ violating sphaleron interactions are simultaneously
in equilibrium with the lepton number violating interactions.
The constraints on $\epsilon_i, \kappa_i$ and $\kappa_i^{\prime}$
follow by demanding that the corresponding interaction be out of
thermal equilibrium when the sphaleron interactions are in
equilibrium, i.e. for $T \geq 100\;  GeV$. The rates for the
L-violating interactions characterised by $\epsilon_i$ are
typically given by $\displaystyle{\Gamma_2 \sim {\epsilon^2 \over T}}$. These
interactions are out of thermal equilibrium for $T > 100 \; GeV$ if
$\displaystyle{\Gamma_2 < 20\; H {T^2 \over M_p}}$ for $T\sim T_C \sim 100\;
GeV$ ($H$ is the Hubble constant). This immediately implies \cite{9},
\be
\epsilon_i \leq 10^{-6}\;  GeV.
\label{e6}
\ene
Likewise, requiring the rates for dimension three interactions
characterised by $\kappa_i, \kappa_i^{\prime} \leq H$
to be less than the expansion rate at $T \sim T_C$ one obtains
\be
\kappa_i, \kappa_i^{\prime} \leq 10^{-4}\;  GeV^2.
\label{e7}
\ene
The constraints on $\kappa_i, \kappa_i^{\prime}$ can be
translated into constraints on the vev $\omega_i$ through eq.
(5). If one takes $v_1 \sim v_2 \sim \mu_{L_i} \sim 100 \;
GeV$ then eq. (7) implies,
\be
\omega_i \leq 10^{-6}  GeV.
\label{e8}
\ene
The exact limits on $\omega_i$ depend upon the model parameters.
But we shall regard limit on $\omega_i$ as given in eq.
(\ref{e8}) as indicative of the typical limit and work out the
consequences of eqs. (6) and (8) in the next section.

The restrictions displayed in eqs. (6) and (8)
are generic constraints which hold in a general situation. If
some of the $\epsilon_i $ and $\kappa_i^{\prime}$ are zero then
the theory would automatically possess a global lepton number
symmetry corresponding to the $i^{th}$ lepton number.  The
presence of such a global symmetry could prevent \cite{17} the erasure of
the baryon asymmetry. The other non-zero $\epsilon_i$ would not
be constrained in this case. We shall disregard this possibility
and assume no global symmetry $L_i$ to be exact.

\section {Neutrino masses in SSM}

Neutrino masses arise in the SSM of the last section through
three different sources. Firstly, the non-zero $\epsilon_i$, directly
induce mixing of the neutrino with Higgsino. Secondly, the vev
$\omega_i$ induced by the presence of $\epsilon_i$ give rise to
mixing between neutrinos and gauginos. These two sources
contribute at tree level. But since the lepton number is
violated, one could radiatively generate the direct majorana
mass term among neutrinos \cite{5,6}. Their strength is also controlled by
the basic parameters $\epsilon_i$ and other $R$ breaking
parameters in eq. (2). We shall assume that the tree
level contribution dominates over the radiatively generated
masses. Since the baryogenesis constrains the tree level mass
very significantly $\leq 10^{-5}\; eV$, it is reasonable to
neglect the radiative contributions and concentrate on the tree
level masses.

In addition to the three neutrinos, the SSM contains
two
gauginos ($\tilde{B}$, $\tilde{W}_3$) and two Higgsinos ($\tilde{H}_1$,
$\tilde{H}_2$). The neutralino mass matrix has the following
form \cite{2,4,15}
in the basis $\chi'^T=(\nu,\; \tilde{B},\; \tilde{W}_3, \; \tilde{H}_1,
\; \tilde{H}_2)$
\be \label{e9}
{\cal M}_0=\left(\begin{array}{cc}
0 & m \\
m^T & M_4 \end{array}\right) \ene
where $m$ is a $3 \times 4$ matrix given by
\be \label{e10}
m=\left(\begin{array}{cccc}
-{g' \over 2}\omega_1 & {g \over 2}\omega_1 & 0 & -\epsilon_1 \\
-{g' \over 2}\omega_2 & {g \over 2}\omega_2 & 0 & -\epsilon_2 \\
-{g' \over 2}\omega_3 & {g \over 2}\omega_3 & 0 & -\epsilon_3
\end{array}\right)
\ene
and $M_4$ is the usual neutralino mass matrix \cite{2} describing
neutralino mixing in the absence of $R$-parity breaking
\be \label{e11}
M_4=\left(\begin{array}{cccc}
cM & 0 & -{1 \over 2}g'v_1 & {1 \over 2}g'v_2 \\
0 & M & {1 \over 2}gv_1 & -{1 \over 2}g'v_2 \\
{1 \over 2}g'v_2 & -{1 \over 2}gv_2 & 0 & - \mu \\
-{1 \over 2}g'v_1 & {1 \over 2}gv_1 & - \mu & 0 \end{array}\right)\ene
$M$ is the common gaugino mass parameter and
$\displaystyle{c={5\alpha_1 \over \alpha_2}=0.5}$ \cite{15}. Since the
parameters ($\epsilon_i$, $\omega_i$) entering $m$ are expected
to be much smaller than the ones appearing in $M_4$, the
neutralino mass matrix ${\cal M}_0$ has a seesaw structure.
Hence the the neutrino masses and mixing are derived from an
effective mass matrix of the form
\be \label{e12}
m_{eff.}=-m\; M_4^{-1}\; m^T = {(cg^2+ g'^2) \over D}
\left(\begin{array}{ccc}
A_1^2 & A_1 A_2 & A_1 A_3 \\
A_1 A_2 & A_2^2 & A_2 A_3 \\
A_1 A_3 & A_2 A_3 & A_3^2 \end{array}\right) \ene
where we have defined
\be \label{e13}
\vec{A}\equiv \mu \vec{\omega} - v_1 \vec{\epsilon} \ene
and
\be \label{e14}
D \equiv 4{det M_4 \over M}=2\mu \left[-2cM\mu + v_1v_2
\left(cg^2 + g'^2 \right)\right] \ene

Although the $7 \times 7$ neutralino mass matrix is quite
complex, the neutrino masses can be approximately described by a
simple structure displayed in eq. (12). $m_{eff.}$ can be
diagonalized by an orthogonal matrix $O$
\be \label{e15}
O\; m_{eff.}\; O^T = diag(0, \; 0, \; m_{\nu}) \ene
with the only non-zero neutrino mass given by
\be \label{e15p}
m_{\nu}=tr(m_{eff.})={(cg^2+ g'^2) \over D}\vert \vec{A} \vert^2
\ene
The matrix $O^T$ can be parametrized by
\be \label{e16}
O^T=\left(\begin{array}{ccc}
\cos \theta_{13} & 0 & -\sin \theta_{13} \\
\sin \theta_{23}\sin \theta_{13} & \cos \theta_{23} & \sin \theta_{23}
\cos \theta_{13} \\
\sin \theta_{13} & \sin \theta_{23} & \cos \theta_{13}\cos \theta_{23}
\end{array}\right) \ene
with the mixing angles given by
\be \label{17}
\tan \theta_{13} = -{A_1 \over \sqrt{A_2^2 + A_3^2}}, \;\;\;\;\;
\tan \theta_{23} = {A_2 \over A_3} \ene
We note that \begin{itemize}
\item two of the eigenvalues of $m_{eff.}$ are zero. This
is not an artifact of the seesaw approximation, but follows in a
more general situation with the full $7 \times 7$ matrix ${\cal
M}_0$. It is easy to see that the following $\psi_0$ represents
two eigenvectors of ${\cal M}_0$ with zero eigenvalues
\be \label{e18}
\psi_0^T = \left((\vec{\epsilon}\wedge \vec{\omega})_1 x_3 -A_2 x_6,\;
(\vec{\epsilon}\wedge \vec{\omega})_2 x_3 +A_1 x_6, \;
(\vec{\epsilon} \wedge \vec{\omega})_3 x_3, \;
0,\; 0, \; x_6, \; 0 \right) \ene
for arbitrary $x_3$ and $x_6$. This feature of the neutralino
masses is a direct consequence of the restricted structure of
${\cal M}_0$ implied by the particle and charge assignments in SSM.

\item The non-zero eigenvalue is typically given by
\be \label{e19}
m_{\nu} \;\; \sim \;\; {\epsilon^2 \over M} \ene
for  $\epsilon \sim 10^{-6}\; GeV$ and $M \sim 100 \; GeV$ one has
$m_{\nu} \sim 10^{-5}\; eV$. This is in the right range for a solution
of the solar neutrino problem through vacuum oscillations\cite{sant}

\item If $\epsilon_i$ do not display any hierarchy then both the mixing angles
 are automatically large. In fact for
$\epsilon_1 \sim \epsilon_2 \sim \epsilon_3$ and $\omega_1 \sim
\omega_2 \sim \omega_3$ we have,
\be \label{e20}
\tan \theta_{23} \sim 1, \;\;\;\;\;\; \tan \theta_{13} \sim -{1
\over \sqrt{2}}
\ene
Thus if all $\epsilon_i$ and $\omega_i$ are flavour independent
and near their limit coming from baryogenesis then one naturally
generates the (mass)$^2$ difference and the mixing angles
required for the vacuum solution to the solar neutrino problem.
The details depend upon other parameters as well and we will now
present the quantitative analysis.
\end{itemize}
Given the mixing matrix $O$  and the mass $m_\nu$, the survival
probability for the solar $\nu_e$ after time $t$ is given by,
\be \label{21}
P_{\nu_e \nu_e}  = 1 - \sin^2\theta_{13} \sin^2{\Delta \cdot t \over
{2 p}}
\ene
where $\Delta \equiv m_\nu^2$. This displays a simple two
generation like structure due to the fact that there is only one
non-trivial (mass)$^2$ difference. The restrictions on the
parameters $\theta_{13}$ and $\Delta$ have been worked out in
detail \cite{12} combining observations of all the four solar neutrino
detectors. The allowed ranges of these parameters are given by
\be \label{22}
\Delta \simeq (0.5 - 1.0) \times 10^{-10}  eV^2, \;\;\;\;\;
\sin^2(2\theta_{13}) \simeq 0.75 - 1.0
\ene
Note that the expected value of $\displaystyle{\tan \theta_{13} \sim -{1 \over
\sqrt{2}}}$ when $\epsilon_i$ and $\omega_i$ are flavour
independent falls within the allowed range. The exact value of
$\Delta$ depends upon the parameters of the SSM in addition to
$\epsilon_i$ and $\omega_i$. These SSM parameters are tightly
constrained by various observations at LEP and  $p \bar p$ collider.
 We shall use these
constraints and show that values of  $\Delta $ in the above
range are possible for $\epsilon_i$ and $\omega_i$
near the limit from baryogenesis.

Various observables in the SSM can be expressed in terms of
the three basic parameters $\mu, M $ and $\tan \beta$. Observations
at LEP and $p \bar p$ collider have been used \cite{17} to
restrict these parameters. In the following we fix the
 $\omega_i, \epsilon_i$ near their limit coming from
baryogenesis and then show that it is possible to obtain the
$\Delta$ in the band required for the vacuum oscillations
solution to  the solar neutrino problem, for a range of values
$(\mu ,M)$ allowed by the other observables. Specifically we
choose $\omega_i = \epsilon_i \equiv\frac{ \omega}{\sqrt 3} $
independent of the
flavour. As already remarked, for these values, the mixing angle
$\displaystyle{\tan \theta_{13} = -{1 \over \sqrt{2}}}$ and is in the
allowed band.

The constraints on $\mu , M $ and $\tan \beta$ coming from   the
non-observation of the decay $ Z \rightarrow \chi^+ \chi^- $,
($ \chi^{\pm}$  being the chargino) is found \cite{18} to be  very
restrictive among various restrictions that are possible
on the SSM parameters from the LEP and $ p \bar p $ collider
experiments \cite{19}. We reproduce in Fig. 1 (solid line)  the
allowed  values of $\mu$ and $M$ for $\tan \beta  = 4 $,
obtained by requiring that the lighter of the charginos be
heavier than $45\;  GeV$. To be specific we have taken $\omega^2 =
2 \times 10^{-12}\; GeV^2$ and plotted the curves in the $\mu ,
M $ plane corresponding to $\Delta = 10^{-10}\; eV^2$ and
$\Delta = 0.5 \times 10^{-10}\; eV^2$ (dotted). It is seen that
there exists a sizable region in the $\mu , M $ plane which is
allowed by various observations and which offers a solution to
the solar neutrino problem. The allowed region is dependent on the
chosen values of $\epsilon$ and $\omega$. This dependence is
displayed in Fig. 2.
which shows the variation of $\Delta$ with the basic parameters $\epsilon_i$
and $\omega_i$ . We once again assume
$\epsilon_i \equiv\frac{ \epsilon}{\sqrt 3}$ and $\omega_i \equiv\frac
{ \omega}{\sqrt 3}$  independent of $i$ and plot the
region in $\epsilon$ and $\displaystyle{\delta \equiv {\epsilon \over \omega}}$
plane, for which $\Delta$  lies between $10^{-10} \;eV^2$ and $5
\times 10^{-11}\; eV^2$. $\tan \beta$ is chosen to be $4$ and the
curves are shown for two typical values of the pair ($\mu , M$). This figure
highlights the fact that for $\epsilon \sim \omega $ and $\epsilon$
in the range allowed by baryogenesis constraint, one could get
$\Delta$ in the range required for solving the solar neutrino problem,

 The bound on the neutrino mass ($m_{\nu}\leq 10^{-5}$ eV) following here
from the baryogenesis constraints
is to be contrasted with
an  analogous bound \cite{20} on the (majorana) masses of neutrinos in
generic seesaw type model. The lepton number violation appears
in this model through large Majorana mass for the neutrino. It
generates the  left-handed neutrino mass through dimension five
operator. Baryogenesis constraints on this operator lead to a
typical mass $m_\nu < 50\; eV$. In contrast,  dimension 2 and 3
terms are responsible for the neutrino masses and the
baryogenesis constraint on them translates into a much stronger
limit $m_\nu < 10^{-5}\; eV$ in the context of the SSM considered
here.


\section{Neutralino Decay}
The lightest R-odd particle is not allowed to decay in the SSM
with R symmetry. Due to its stability and neutrality, the LSP is considered
an ideal candidate for the dark matter of the universe
\cite{dm}. Usually, one expects a combination of neutralinos to
be the LSP \cite{2}. The presence of even a tiny amount of R violation
such as the one considered here can make the LSP unstable on
cosmological scale. The couplings which make the LSP unstable
have been considered in the literature \cite{4,8,15}.
 The  treatment
of the neutralino mass matrix based on the seesaw approximation
makes it possible to write down the couplings of neutralino to
neutrino analytically without neglecting the inter generational mixing.
 We give this couplngs below for completeness and discuss their
consequences.

R parity violation causes mixing between two neutral particles
(neutrinos and gauginos ) transforming differently under the
$SU(2)\times U(1)$ group. As is well known, the couplings of Z
to fermions no longer remain flavour diagonal in this case.
Neutralinos couple to Z through the following equation in the weak
basis $\chi'$:
\begin{equation}
-{\cal L_{Z}}=\frac{g}{2\cos\theta_{W}}\left[ {\overline\chi}'\;
                           \overline{\sigma}_{\mu} T_3\; \chi'
                                                    \right]\;Z^{\mu}
\end{equation}
where $T_3\equiv (1,1,1,0,0,-1,1)^T$.
Let $U$ be the matrix which diagonalizes the ${\cal M}_0$ of eq.(1)
\begin{equation}
U\;{\cal M}_0\;U^T=diag. (0,0,m_{\nu},m_4,m_5,m_5,m_7)
\end{equation}
$m_{\alpha}, \;\; \alpha=4,5,6,7$ are neutralino masses.
The form of $U$ is well known in the seesaw approximation
\begin{equation}
U=\left(
       \begin{array}{cc}
        O_{\nu}(1-\frac{1}{2}\rho \rho^T)&-O_{\nu}\rho\\
        O_{\lambda}\rho^T&(1-O_{\lambda}\frac{1}{2}\rho^T \rho)\\
        \end{array}  \right)
\end{equation}
Here $\rho\equiv m\;M_4^{-1}$. $O_{\nu}(O_{\lambda})$  diagonalize
$m_{eff} (M_4)$ of eqs.(11,12).

The flavour non diagonal coupling of neutralinos to Z follow from
eqs.(24-26) after some algebra:
\begin{equation}
-{\cal L}_{\nu\;\chi}=\frac{g}{4\cos\theta_{W}}\left[
         F^{*}_{i\alpha} {\overline
\nu}_{iL}\gamma_{\mu}\chi_{\alpha L}
         -F_{i\alpha} {\overline \nu}_{i R}\gamma_{\mu}\chi_{\alpha R }
                                                  \right] Z^{\mu}
\end{equation}
where
\begin{equation}
F_{i\alpha}=-\frac{-2\mu \vert \vec{A} \vert}{D}
(g'(O_{\lambda})_{\alpha 4}-c g (O_{\lambda})_{\alpha 5}) \delta_{i3}
-[\frac{v_2}{\mu}( g'^2+c g^2)(O_{\nu})_{ij}A_j'-
             4 c \epsilon_j M](O_\lambda)_{\alpha 6}
\end{equation}
with $A'_i=A_i+2 \epsilon_i v_1$ and $\vert \vec{A} \vert$ and $D$
are defined before.

The elements $O_{\alpha \beta}$ appearing in  equation above
depends upon the composition of the LSP in terms of
neutralinos. They are given as in the MSSM \cite{2}.
But it follows from the structure of the above equation that
couplings of the LSP to $Z \nu$ are  not suppressed by mixing factor coming
from the neutralino mixing \cite{fn}. Thus irrespective of its composition,
the LSP will decay into a real (or virtual) $Z$ and the neutrino.
The typical strength is given from eq.(28) by $\frac{g\epsilon}{\mu}$.
For $\epsilon \sim 10^{-6}$ GeV, this strength is strong enough
to make the LSP decay on the cosmological scale but is  not
large enough to cause its decay and hence any signature in the laboratory.
 The life time following from eqs
.(27,28) is typically given by
\begin{eqnarray}
\tau&\sim& \tau_{\mu} \left(\frac{g\epsilon}
{\mu}\right)^{-2}\;\;\left (\frac{M_{LSP}}{m_{\mu}}\right)^{-5} \nonumber \\
&\sim & 1.6 \times 10^{-4}\;\; {\rm sec}\;\;\left (\frac{M_{LSP}}{50
 GeV}\right)^{-5}
 \;\;\left  (\frac{
100 GeV}{\mu}\right)^{-2}\;\;\left (\frac{\epsilon}{10^{-6} GeV}\right)^{-2}\\
\end{eqnarray}
where $\tau_{\mu}$ is the muon lifetime. It is seen from above that
the typical LSP with 50 GeV mass will not be able to decay inside the detector
but it will be short lived to have any cosmological signature.
The LSP in this case would contribute to the invisible Z width but
this contribution is suppressed by the factor $(\frac{\epsilon}{\mu})^2$
compared to other fermion contributions and thus is practically negligible.

\section{Conclusions}

We have discussed in this paper possibility of obtaining a
solution to the solar neutrino problem through vacuum
oscillations in the SSM. This requires extremely tiny $({\rm
mass})^2$ difference $\Delta \sim 10^{-10}\; eV^2 $.
As we have discussed, the limit on the neutrino masses coming
from baryogenesis imply $\Delta \leq 10^{-10}\;eV^2$. Hence if
$R$-breaking parameters are near this limit then SSM offers a
vacuum solution to the solar problem. This becomes more
interesting due to the fact that the relavant mixing angle predicted in
this model is naturally large as is required for the vacuum
solution.

A similar analysis of MSW solution to the solar neutrino problem
in supersymmetric theory has been earlier carried out in an
extension of SSM involving right-handed neutrinos and other
gauged singlet superfields \cite{7}. This analysis
concentrated on the spontaneous violation of $R$ rather than the explicit.
In such a case the baryogenesis may not restrict the amount of $R$ violation.
In contrast,
it is not possible to break $R$ spontaneously  in the minimal case considered
 here,
and one should
then consider baryogenesis
constraints. In spite of its strongness, these constraints do allow
interesting physical effect namely a solution to the solar neutrino problem as
 we
we have argued here.

It is hard to explain why the $R$-parity breaking
parameters $\epsilon_i$ are  as small as is required from
baryogenesis limit. But if they do have such values then they
may be responsible to cause vacuum oscillations of the solar neutrinos.

\section{ Acknowledgments}

We thank S.Rindani and R.M.Godbole for valuable discussions.
A.S.J. wants to thank A. Masiero for helpful discussions.
Thanks are also due to J. W. F. Valle and B. Ananthanarayan for
critical comments on the manuscript.
M.N. wishes to thank the Alexander von Humboldt foundation for
financial support under the Feodor-Lynen Fellowship program.
The work of A.S.J was supported by the DGYCT grant SAB94-0014.
\newpage
\noindent {\bf Figure Caption}\\

\noindent {\bf Fig. 1}. The allowed region in the $\mu-M$ plane
corresponding to the chargino  mass $< 45 \;GeV $(solid) and
$5\times 10^{-11}\;<\Delta\;<10^{-10} eV^2$ (broken). $\tan \beta$ is
chosen to be 4 and $\omega^2=2 \times 10^{-12} eV^2$. The region above each
 curve is allowed.\\[.5cm]
\noindent {\bf Fig. 2}. Band of values of $\omega$ and $\frac{\epsilon}
{\omega}$ allowed by $5\times 10^{-11}\;<\Delta\;<10^{-10} eV^2$.
The solid and the dotted curves are  for $(\mu,M)$=(100,100) GeV and (-100,100)
 GeV
respectively. $\tan \beta$ is chosen to be 4. The region above each
 curve is allowed.


\newpage
\begin{thebibliography}{99}
\bibitem{one}
A.~Yu.~Smirnov, ICTP preprint IC/93/388; S.~M.~Bilenky and
S.~T.~Petkov, Rev.~Mod.~Phys.~{\bf 59} (1987) 671; J.~W.~F.~Valle,
Pro.~Part.~Nucl.~Phys.~{\bf 26} (1991) 91.

\bibitem{2}
H.~E.~Haber and G.~L.~ Kane, Phys.~Rep.~{\bf 117}
(1985) 75; for recent review see X.~Tata in {\it The Standard
Model and Beyond} p.~304, ed.~J.~E.~Kim, World Scientific 1991.

\bibitem{3}
S.~Weinberg, Phys.~Rev.~{\bf D26} (1982) 533; N.~Sakai and
T.~Yanagida, Nucl.~Phys.~{\bf B197} (1982) 533;
Recent review can be found in D.~P.~Roy, invited talk at the $X$
Symposium in High Energy Physics, Bombay, TIFR preprint TIFR/TH/93-14.

\bibitem{4}
J.~Ellis et al., Phys.~Lett.~{\bf B150} (1985) 142.

\bibitem{5}
L.~Hall and M.~Suzuki, Nucl.~Phys.~{\bf B231} (1984) 419; I.~Lee
Nucl.~Phys.~{\bf B246} (1984) 419.

\bibitem{6}
K.~Enqvist, A.~Masiero and A.~Riotto, Nucl.~Phys~{\bf B373}
(1992) 95;E. Roulet and D. Tommasini, Phys. Lett. {\bf B256} (1991) 218;
 D. Tommasini, Phys. Lett. {\bf B297} (1992) 125.

\bibitem{7}
J.C. Romao and J. W. F. Valle, Phys. Lett. {\bf B272} (1991) 436; Nucl. Phys.
 {\bf B 381} (1992) 87.

\bibitem{8}
R.~M.~Godbole, P.~Roy and X.~Tata, Nucl.~Phys.~{\bf B401} (1993)
67

\bibitem{9}
B.~Campbell et al., Phys.~Lett.~{\bf B256} (1991) 457; Astroparticle Physics,
{\bf 1} (1992) 77; W. Fishler, G. F. Giudice, R. G. Leigh and S. Paban,
Phys. Lett. {\bf B258} (1991) 45.

\bibitem{10}
For a recent review and further references see A.~G.~Cohen,
D.~B.~Kaplan and A.~E.~Nelson,
Annu.~Rev.~Nucl.~Part.~Sci.~{\bf 43} (1993) 27.

\bibitem{11}
S.~P.~Mikheyev and A.~Yu.~Smirnov, Sov.~J.~Nucl.~Phys.~{\bf 42}
(1985) 913; JETP {\bf 64} (1986) 4; L.~Wolfenstein,
Phys.~Rev.~{\bf D17} (1978) 2369

\bibitem{12}
S.~L.~Glashow and L.~M.~Krauss, Phys.~Lett.~{\bf B190} (1987) 199;
V.~Barger, R.~J.~N.~Phillips and K.~Whisnant, Phys.~Rev.~{\bf
D43} (1991) 1110; A.~Acker, S.~Pakvasa and J.~Pantaleone, Phys.~Rev.~
{\bf D43} (1991) 1754. A recent analysis can be found in P.I. Krastev and
S.T. Petcov,Phys. Lett. {\bf B285} (1992) 85.

\bibitem{Planck}
E.Kh. Akhmedov,Z.G. Berezhiani and G. Senjanovick, Phys. ReV. Lett.
{\bf 69} (1992) 3013.

\bibitem{13}
C.~S.~Aulakh and R.~N.~Mohpatra, Phys.~Lett.~{\bf B190} (1982) 136.

\bibitem{14} G.~G.~Ross and J.~W.~F.~Valle, Phys.~Lett.~{\bf
B151} (1985) 375.

\bibitem{15}
D.~Comelli et al., SISSA-PREPRINT, SISSA 93/168-A.

\bibitem{16}
R.~Barbieri et al., Phys.~Lett.~{\bf B238} (1990) 86;
D. E. Brahm et al., Phys. ReV. {\bf D42} (1992) 1860.

\bibitem{asj}
A.~S.~Joshipura and M.~Nowakowski, PRL-TH-94/11.

\bibitem{17}
A.~Nelson and S.~M.~Barr, Phys.~Lett.~{\bf B246} (1990) 141.

\bibitem{sant}
Note that the sneutrino vev is constrained to be $< 10 \; keV$ from stellar
cooling in models with spontaneous violation of R parity.
 The neutrino masses fall in this case  in the range appropriate
for the MSW solution to the solar neutrino problem. See for example,
A. Santamaria and J.W.F. Valle, Phys.~Lett.~{\bf B195} (1987) 423.
This scenario
is however not viable in view of the LEP constraints on the invisible width.


\bibitem{18}
A recent analysis and original references can be found in
G.~Giacommeli and P.~Giacommeli, CERN-PPE/93-107. Earlier
analysis is contained in J.~Ellis, G.~Ridolfi and F.~Zwirner,
Phys.~Lett.~{\bf B237} (1990) 423.

\bibitem{19}
The non-observation of neutralino puts a somewhat more stringent
restrictions than that of the chargino displayed in Figs. 1,2. See,
e.g., [18].

\bibitem{20}
M.~Fukugita and T.~Ynagida, Phys.~Rev.~{\bf D42} (1990) 1285.

\bibitem{dm}
See for example, Lopez, J. L., Nanopoulos D. and Yuan K., Nucl. Phys.
 {\bf B370} (1992) 445.

\bibitem{fn}
 This follows from the fact that all  $O_{\alpha 4},O_{\alpha 5},O_{\alpha 6}$
appear in eq.(28) and the LSP is never a pure $\tilde{H}_2$, see
\cite{2}.
\end{thebibliography}
\end{document}